\documentclass[twocolumn,prl,superscriptaddress,floatfix]{revtex4-2}

\usepackage{graphicx}
\usepackage{xcolor}
\usepackage{epsfig}
\usepackage{rotating}
\usepackage{amsmath}
\usepackage{amsfonts}
\usepackage{amssymb}
\usepackage{enumerate}
\usepackage{longtable}
\usepackage{appendix}
\usepackage{hyperref}
\usepackage{dcolumn} % Align table columns on decimal point
\usepackage{bm}
\DeclareMathOperator{\Tr}{Tr}
\definecolor{nicered}{RGB}{210,50,0}
\usepackage{hyperref}\hypersetup{
    colorlinks=true,
    linkcolor=nicered,
    filecolor=magenta,      
    urlcolor=magenta,
    citecolor=blue
    }
\usepackage{ulem}
\usepackage{braket}

\begin{document}

\title{High-temperature magnetization and entropy of the triangular lattice Hubbard model in a Zeeman field}

\author{Owen Bradley}
\affiliation{Department of Physics, University of California Davis, California 95616, USA}

\author{Yutan Zhang}
\affiliation{Department of Physics, University of California Davis, California 95616, USA}

%\author{Pranav Seetharaman}
%\affiliation{Department of Physics, San Jose State University, San Jose, California 95192, USA}

%\author{Ehsan Khatami}
%\affiliation{Department of Physics, San Jose State University, San Jose, California 95192, USA}

\author{Jaan Oitmaa}
 \affiliation{School of Physics, The University of New South Wales, Sydney 2052, Australia}

\author{Rajiv R. P. Singh}
 \affiliation{Department of Physics, University of California Davis, California 95616, USA}

\date{\rm\today}

\begin{abstract}
    We use strong coupling expansions to calculate the entropy function $S(T,h)$, the magnetization $M(T,h)$, and the double occupancy factor $D(T,h)$ for the half-filled triangular lattice Hubbard model as a function of temperature $T$ and Zeeman field $h$, for various values of the Hubbard parameter ratio $U/t$. These calculations converge well for temperatures larger than the exchange parameter $J=\frac{4t^2}{U}$ for moderate to large $U/t$ values. Setting $\mu=U/2$ suffices to obtain the density of half filling within a fraction of one percent at all temperatures studied for $U/t \geq 8$.
    %We verify that the Maxwell relation between $\partial S/ \partial h$ and $\partial M/ \partial T$ is satisfied. 
    We discuss the systematic variation of properties with $U/t$. The temperature dependence of 
    entropy and the double occupancy parameter shows a mapping to an antiferromagnetic Mott insulating behavior at temperatures well above $T=J$ for $U/t\ge 10$. 
   %% A comparison with the Heisenberg model allows us to study the crossover to the Heisenberg behavior as double occupancy is suppressed. 
   % The zero-field magnetic susceptibility shows a crossover from the Heisenberg behavior at low temperatures to one where the Curie constant is reduced by a factor of $2$ at temperatures of order $U$.
    Convergence of the series is weaker at intermediate fields implying non-monotonic 
    variation of spin-correlations with the Zeeman field. 
    We discuss the relevance of the Hubbard model results to the triangular-lattice antiferromagnetic materials Lu$_3$Cu$_2$Sb$_3$O$_{14} $ (LCSO) 
    %and Lu$_3$CuZnSb$_3$O$_{14}$ (LCZSO) 
    studied recently by Yang {\it et al }[Yang et al arXiv:2102.09271 (2022)].    
    % The magnetization and entropy show hints of magnetization plateaus up to surprisingly high temperatures.
\end{abstract}

\maketitle

\section{I. Introduction}

The Hubbard model serves as one of the most important paradigms for studying electronic correlations in solids \cite{Hubbard1963,Kanamori1963,Gutziwiller1963}. It has played a central role in our emerging understanding of many important solid-state phenomena \cite{Georges1996,Fradkin2015,LeBlanc2015} including metal-insulator transitions, quantum magnetism, the physics of intertwined orders, strange metals, and high temperature superconductivity. In recent years, the model has found additional relevance in new platforms such as cold-atoms in optical lattices \cite{Duarte2015,Chiu2018,Hofrichter2016} and few layers of Van der Waals systems such as graphene and transition-metal dichalcogenides \cite{Wu2018} that can be manipulated to produce a plethora of exotic phases. Advantages of some of these new platforms over traditional solid state systems are that the Hubbard model parameters can be controlled and well-determined, and the system can be relatively free from unknown perturbations and engineered to exquisite detail. 

The Hubbard Hamiltonian is given by
\begin{align}
    {\cal H} &= -t \sum_{\langle \mathbf{i}, \mathbf{j} \rangle, \sigma} \left( c^\dagger_{\mathbf{i} \sigma} c_{\mathbf{j} \sigma}  + c^\dagger_{\mathbf{j} \sigma} c_{\mathbf{i} \sigma} \right) + U \sum_{\mathbf{i}} n_{\mathbf{i} \uparrow} n_{\mathbf{i} \downarrow} \label{Ham} \\
    &- \mu \sum_{\mathbf{i}} \left( n_{\mathbf{i} \uparrow} + n_{\mathbf{i} \downarrow} \right) - \frac{h}{2} \sum_{\mathbf{i}} \left( n_{\mathbf{i} \uparrow} - n_{\mathbf{i} \downarrow} \right) \nonumber
\end{align}
where $c^\dagger_{\mathbf{i} \sigma} (c_{\mathbf{i} \sigma})$ is a creation (destruction) operator for an electron with spin $\sigma=\{ \uparrow, \downarrow \}$ at site $\mathbf{i}$ of a lattice. Here $t$ is the nearest-neighbor hopping parameter, and the first sum is taken over all nearest neighbor pairs of sites $\langle \mathbf{i}, \mathbf{j} \rangle$. The number operator $n_{\mathbf{i} \sigma} = c^\dagger_{\mathbf{i}\sigma}c_{\mathbf{i}\sigma}$ gives the number of electrons at site $\mathbf{i}$ of spin $\sigma$. An on-site repulsion $U>0$ penalizes sites which are doubly occupied, while the chemical potential $\mu$ controls the overall filling. 
%For bipartite lattices, $\mu=U/2$ corresponds to half-filling, i.e.~an average occupation of one electron per site. 
We include a Zeeman field $h$ which lowers the overall energy for electron spins oriented in the direction of the magnetic field.

In this paper we wish to explore properties of the half-filled triangular-lattice Hubbard model at temperatures above the exchange energy scale $J$. There have been many experimental studies of triangular antiferromagnets over the last few years \cite{Susuki2013,Yokota2014,Rawl2017,Paddison2017,Cui2018,Fortune2021,Okada2022} reporting varied phase behaviors including different types of quantum spin-liquids, yet many questions remain. We are motivated, in part, by the recent experimental study of the triangular-lattice antiferromagnetic materials Lu$_3$Cu$_2$Sb$_3$O$_{14} $ (LCSO)
and Lu$_3$CuZnSb$_3$O$_{14}$ (LCZSO) by Yang et al \cite{Yang2022}. These materials have an exchange energy scale
$J$ of order $10 K$. However, the authors find that the molar magnetic entropy difference $\Delta S (T)$ between 
temperatures of $T=0.1 K$ and above $T=20K$, where it begins to saturate as a function of temperature, was only about a third of $R\ln{2}$. This is an extraordinary result implying either that at temperatures an order of magnitude below $J$ a substantial fraction of $R\ln{2}$ entropy remains in
the system or that the entropy at $T=2J$ is already reduced to only a third of $R\ln{2}$. 

From a theoretical point of view, the nearest-neighbor Heisenberg model on the triangular-lattice is known to retain
substantially more entropy as temperature is lowered compared to the square-lattice. Using high temperature series expansions for the Heisenberg model, Elstner {\it et al} found \cite{Elstner1993,Elstner1994} that the square-lattice entropy at $T=0.4J$ was about ten percent of $R\ln{2}$ for the square-lattice but close to fifty percent of $R\ln{2}$ for the triangular lattice. However, at the lowest temperatures the triangular-lattice entropy is decreasing with reduction in temperature and one does not expect much entropy to remain \cite{Bernu2001,Kulagin2013} at $T=0.1 J$. Indeed, there is overwhelming evidence that the triangular-lattice Heisenberg model has an
ordered ground state \cite{Elser1988,Singh1992,Bernu1994,Capriotti1999,White2007,Chen2013,Zhu2018,Oitmaa2020} and extrapolations of high temperature series that build in information about the
low temperature behavior imply \cite{Bernu2001} a very small entropy at $T=0.1 J$. 

Here we will study how the magnetic and thermodynamic properties of a finite-$U$ Hubbard model differ from the Heisenberg model at temperatures of order $J$ and
higher and if the behavior of LCSO can be explained by such a finite $U$.
To study the Hubbard model, we will use strong coupling expansions carried out at arbitrary temperatures \cite{Oitmaa1992,Oitmaa2006,Singh2022}.
These are expansions around the atomic limit in powers of $\beta t$. The Hubbard parameter $U$ is treated
non-perturbatively and enters as $\exp{(-\beta U)}$ and as energy denominators $1/U$. These expansions
are very accurate at temperatures above $t$ and turn into an expansion in powers of $t/U$ or $\beta t^2/U$
at lower temperatures. As $T\to 0$, these expansions turn into degenerate perturbation theory around the atomic limit \cite{MacDonald1988,Yang2010}.
We will study the crossover from high-temperatures comparable to $U$ to the strongly correlated regime at temperatures much below $U$. By studying the entropy and magnetization as a function of temperature, Zeeman field $h$, and $U/t$ ratio, we will explore the crossover to Heisenberg behavior and deviations from the Heisenberg behavior ultimately indicating a transition away from a Mott insulator. Note that we use Numerical Linked Cluster (NLC) expansions \cite{Rigol2006,Rigol2007} to obtain results for the triangular-lattice Heisenberg model.

Much of this physics has been studied before at low temperatures \cite{MacDonald1988,Yang2010}. However, it is difficult to establish 
the low temperature properties of the model in a conclusive way for lack of definitive computational tools. Our goal is to understand the intermediate temperature properties of the model where the strong coupling expansions should be well converged and provide accurate answers in the thermodynamic limit.

We also study the Maxwell's relations between $(\frac{\partial S} {\partial h})_T$ and $(\frac{\partial M} {\partial T})_h$. These are important for experimental measurements of magnetic entropy and heat capacity of solid state systems \cite{Yang2022}. Direct experimental measurement of thermodynamic properties such as magnetic entropy and heat capacity can be difficult due to the presence of phonons and other degrees of freedom whose contributions can be hard to accurately subtract. In contrast, the magnetization measurements can be relatively free of such complications. Thus, Maxwell's relations provide a way of measuring field dependent changes in entropy. In a theoretical model study these relations also act as checks on the numerical convergence of the calculations.

We find that for $U/t\ge 16$, the zero-field molar entropy function develops a plateau as a function of temperature at a value of $R\ln{2}$ and below that temperature the Heisenberg behavior is realized. In the
intermediate coupling region $10< U/t<16$, we also see a crossover to Heisenberg behavior but there is no
entropy plateau at a value of $R\ln{2}$ and instead the Heisenberg behavior only sets in at an entropy value 
less than $R \ln{2}$. 
Below $U/t=8$ any resemblance to the Heisenberg behavior is lost. The study
of the double occupancy parameter reinforces the result that a transition away from antiferromagnetic Mott insulator occurs in the region $8<U/t<10$, consistent with previous results \cite{Yang2010,Garwood2022,Mongkolkiattichai2022}. At still smaller $U/t$ the entropy at a temperature of $2J$ is greater than $R \ln{2}$ as the
system still has significant double occupancy left. Around and below $U/t=10$, the entropy decreases rapidly
with reduction in temperature around $T=2J$ and thus smaller $U/t$ values cannot explain the magnetic properties
of these materials at all. In fact, we find that the best fits to the material behavior at high temperatures is obtained 
in the large-$U$ limit, that is by a Heisenberg model.

In our study, we find that one needs a magnetic field of order $J$ to see substantial reduction in the magnetic entropy at temperatures of $2J$ and higher. As expected, strong coupling expansions converge well for high fields
at all temperatures. However, surprisingly, we find that the convergence is worse for intermediate
fields $h\sim 2J$ than at $h=0$. The extrapolations show some hints of magnetization plateaus already 
at temperatures of order $J$, though the convergence remains poor. The non-monotonic convergence suggests
that spin-spin correlation length at these temperatures may be non-monotonic as a function of magnetic field, being
larger at fields of order $2J$, where the system may be developing spatial correlations associated with the onset of magnetization plateaus at lower temperatures.

The organization of the paper is as follows: In Section \hyperref[sec:methods]{II}, we discuss the strong coupling expansion method. In Section \hyperref[sec:resultsA]{III}, the numerical results for zero-field are presented and discussed. In Section \hyperref[sec:resultsB]{IV}, we discuss the thermodynamic properties of the model as a function of magnetic field. In Section \hyperref[sec:LCSO]{V}, we present comparisons of numerical results with the LCSO materials.
In Section \hyperref[sec:conclusions]{VI}, we present our conclusions.

\section{II. Methods}
\label{sec:methods}

We employ finite temperature strong coupling expansions to obtain thermodynamic properties of the triangular-lattice Hubbard model. We write the Hubbard Hamiltonian Eq.~\eqref{Ham} as $H=H_0 + V$, where
\begin{equation}
H_0 = U \sum_{\mathbf{i}} n_{\mathbf{i} \uparrow} n_{\mathbf{i} \downarrow} 
    - \mu \sum_{\mathbf{i}} \left( n_{\mathbf{i} \uparrow} + n_{\mathbf{i} \downarrow} \right) - \frac{h}{2} \sum_{\mathbf{i}} \left( n_{\mathbf{i} \uparrow} - n_{\mathbf{i} \downarrow} \right)
\end{equation}
i.e.~the onsite terms form the unperturbed part of the Hamiltonian, and we treat the electron kinetic energy as a perturbation
\begin{equation}
V = -t \sum_{\langle \mathbf{i}, \mathbf{j} \rangle, \sigma} \left( c^\dagger_{\mathbf{i} \sigma} c_{\mathbf{j} \sigma}  + c^\dagger_{\mathbf{j} \sigma} c_{\mathbf{i} \sigma} \right).
\end{equation}
Now, the logarithm of the grand partition function $\ln Z$ (per site) can be expressed as a perturbation expansion. Using the formalism of thermodynamic perturbation theory \cite{Oitmaa2006}, we expand $\ln Z$ as
\begin{align}
&\ln Z = \ln z_0 + \nonumber \\ 
& \sum_{r=1}^{\infty} \int_0^\beta d\tau_1 \int_0^{\tau_1} d\tau_2 \ldots \int_0^{\tau_{r-1}} d\tau_r \langle \tilde{V}(\tau_1) \tilde{V}(\tau_2) \ldots \tilde{V}(\tau_r) \rangle \label{expansion},
\end{align}
where $z_0$ is the single-site partition function, $\tilde{V}(\tau) = e^{\tau H_0} V e^{-\tau H_0}$, and the expectation value is defined as
\begin{equation}
\langle X \rangle = \frac{\Tr e^{-\beta H_0} X}{\Tr e^{-\beta H_0}}.
\end{equation}
To each order, terms in Eq.~\eqref{expansion} can be expressed in terms of clusters of sites (graphs) on the triangular lattice having $N_s$ sites and $N_r$ bonds. We have that
\begin{equation}
\ln Z = \ln z_0 + \sum_G L_G z_0^{N_s} (\beta t)^{N_r} X_G(\zeta, \beta U) \label{lnZeq},
\end{equation}
where the sum is over graphs $G$. Here $L_G$, the lattice constant for a graph $G$, is the extensive part of its count divided by number of sites, $X_G$ is the reduced weight of the graph, and $\zeta = e^{\beta \mu}$ defines the fugacity of the system.  

The single-site partition function can be expressed as a series in powers of $\zeta$ and $w=e^{-\beta U}$. For the $\mathrm{SU}(2)$ Hubbard model (in a magnetic field $h$) it is given by
\begin{equation}
z_0 = 1 + \left(e^{\beta h/2} + e^{-\beta h/2}\right)\zeta + \zeta^2 e^{-\beta U}.
\end{equation}
We obtain strong coupling expansions for $\ln Z$ up to $8^{th}$ order in $\beta t$. We can rewrite Eq.~\eqref{lnZeq} as a series expansion in orders $\beta t$, i.e.~
\begin{equation}
\ln Z = \ln z_0 + \sum_{n=2}^{\infty} \frac{(\beta t)^n}{z_0^n} \sum_{t=1}^{n_t} a_t(n_1, n_2, n_3, n_4, C) \label{lnZeq2}, 
\end{equation}
where each order $n$ contributes a sum of $n_t$ terms, and our calculations are complete to order $n_{\mathrm{max}}=8$. The function $a_t$ has the form
\begin{equation}
a_t = C e^{\left(n_2 \mu + n_3 h/2 - n_4 U\right)} \left(\beta U\right)^{-(n - n_1)},
\end{equation}
where $n_1$, $n_2$, $n_3$, and $n_4$ take integer values, and the set of $C$ values are the series coefficients that we have calculated complete to eighth order. Without loss of generality, we set the hopping parameter $t=1$.

Thermodynamic properties such as the internal energy $E$, electron density $\rho$, magnetization $M$, double occupancy $D$, and entropy $S$, can now be obtained by taking suitable derivatives of $\ln Z$ given by Eq.~\eqref{lnZeq2}. We have that 
\begin{align}
E &= -\left(\frac{\partial}{\partial \beta} \ln Z \right)_\zeta, \\
\rho &= \zeta \frac{\partial}{\partial \zeta} \ln Z \equiv \frac{1}{\beta}\frac{\partial}{\partial \mu} \ln Z, \\
M &= \frac{1}{\beta} \frac{\partial}{\partial h} \ln Z, \\
D &= -\frac{1}{\beta} \frac{\partial}{\partial U} \ln Z, \label{densityEq}\\
S &= \beta E + \ln Z - \beta \rho \mu, \label{entropyEq}
\end{align}
where each thermodynamic property above is a function of the four parameters $\beta$, $\mu$, $h$, and $U$. In the accompanying Supplemental Material, we provide the non-zero coefficients $C$ for each $n_1$, $n_2$, $n_3$, $n_4$ required to calculate these quantities to $8^{th}$ order. By directly differentiating Eq.~\eqref{lnZeq2} we obtain:
\begin{align}
&E(\beta, \mu, h, U) = e_0 \nonumber \\
&- \sum_{n=2}^{n_\textrm{max}} \sum_{t=1}^{n_t} \frac{a_t \beta^n}{Z_0^n} \left[ \frac{n_3 h}{2} - n_4 U - \frac{(n-n_1)}{\beta} + \left(\frac{n}{\beta} + ne_0\right) \right],\\
&\rho(\beta, \mu, h, U) = \rho_0 + \sum_{n=2}^{n_\textrm{max}} \sum_{t=1}^{n_t} \left[ \left( \frac{\beta^n}{Z_0^n}\right)(n_2 - n\rho_0)a_t \right],\\
&M(\beta, \mu, h, U) = m_0 + \sum_{n=2}^{n_\textrm{max}}\sum_{t=1}^{n_t} \left[ \left( \frac{\beta^n}{Z_0^n}\right) \left(\frac{n_3}{2} - n m_0\right) a_t \right], \\
&D(\beta, \mu, h, U) = d_0 \nonumber \\
&- \sum_{n=2}^{n_\textrm{max}}\sum_{t=1}^{n_t} \left[ \left( \frac{\beta^n}{Z_0^n}\right) \left(n d_0 - n_4 - \frac{(n-n_1)}{\beta U}\right)a_t\right],
\end{align}
where each expression involves a zeroth order term arising from a derivative of $\ln z_0$: $e_0 = -\frac{\partial}{\partial \beta}(\ln z_0)_\zeta$, $\rho_0 = \frac{1}{\beta}\frac{\partial}{\partial \mu}(\ln z_0)$, $m_0=\frac{1}{\beta}\frac{\partial}{\partial h}(\ln z_0)$, and $d_0=-\frac{1}{\beta}\frac{\partial}{\partial U}(\ln z_0)$. Note that the entropy $S$ is subsequently obtained via the thermodynamic relation Eq.~\eqref{entropyEq}.

\section{III. Thermodynamic properties in zero magnetic field}
\label{sec:resultsA}
\begin{figure}[b!]
    \includegraphics[width=\columnwidth]{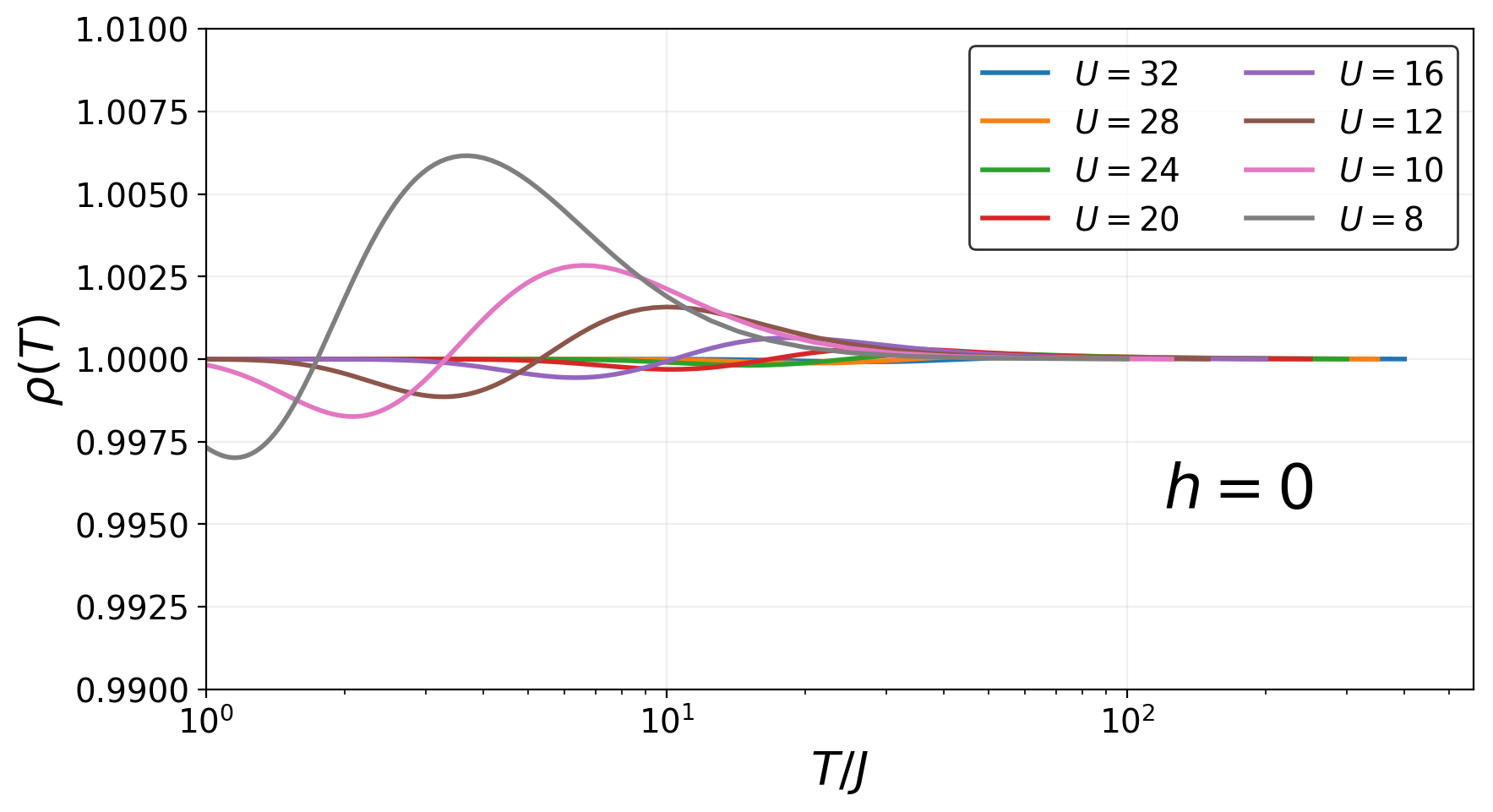}
    \caption{Electron density per site $\rho(T)$ at $h=0$ is shown for different values of $U$, where we fix the chemical potential $\mu = U/2$ in each case. We find this yields half-filling on the triangular lattice to sufficient accuracy (a fraction of one percent) for $U \geq 8$.}
    \label{rho_vs_T}
\end{figure}

We first present results for the triangular lattice Hubbard model in zero magnetic field ($h=0$). On bipartite lattices, fixing the chemical potential at $\mu = U/2$ ensures the system is at half-filling, i.e.~an average electron density of $\rho=1$ per site. However, this does not strictly hold for the triangular lattice, which is non-bipartite. Nevertheless, we find that fixing $\mu = U/2$ yields approximately $\rho=1$ to sufficient accuracy across a wide range of temperatures and values of $U$. In Fig.~\ref{rho_vs_T}, we plot $\rho(T)$ for several values of $U$ from $U=8$ to $U=32$, where we fix $\mu=U/2$, and we report the temperature $T$ in units of the exchange parameter $J=4t^2/U$. We find that setting $\mu=U/2$ suffices to obtain half-filling to within a fraction of one percent at all temperatures studied for $U \geq 8$. We thus fix $\mu = U/2$ in the results that follow. We note that descending from high $T$, there is a slight positive contribution to $\rho$, exceeding half-filling, while at low $T$ there is a small negative contribution yielding $\rho<1$. This behavior can be understood by the examining the third order term in the density series [Eq.~\eqref{densityEq}].
%analytically, as shown in Fig.~A1 (note that $\rho_0=1$ ).

\begin{figure}[t!]
    \includegraphics[width=\columnwidth]{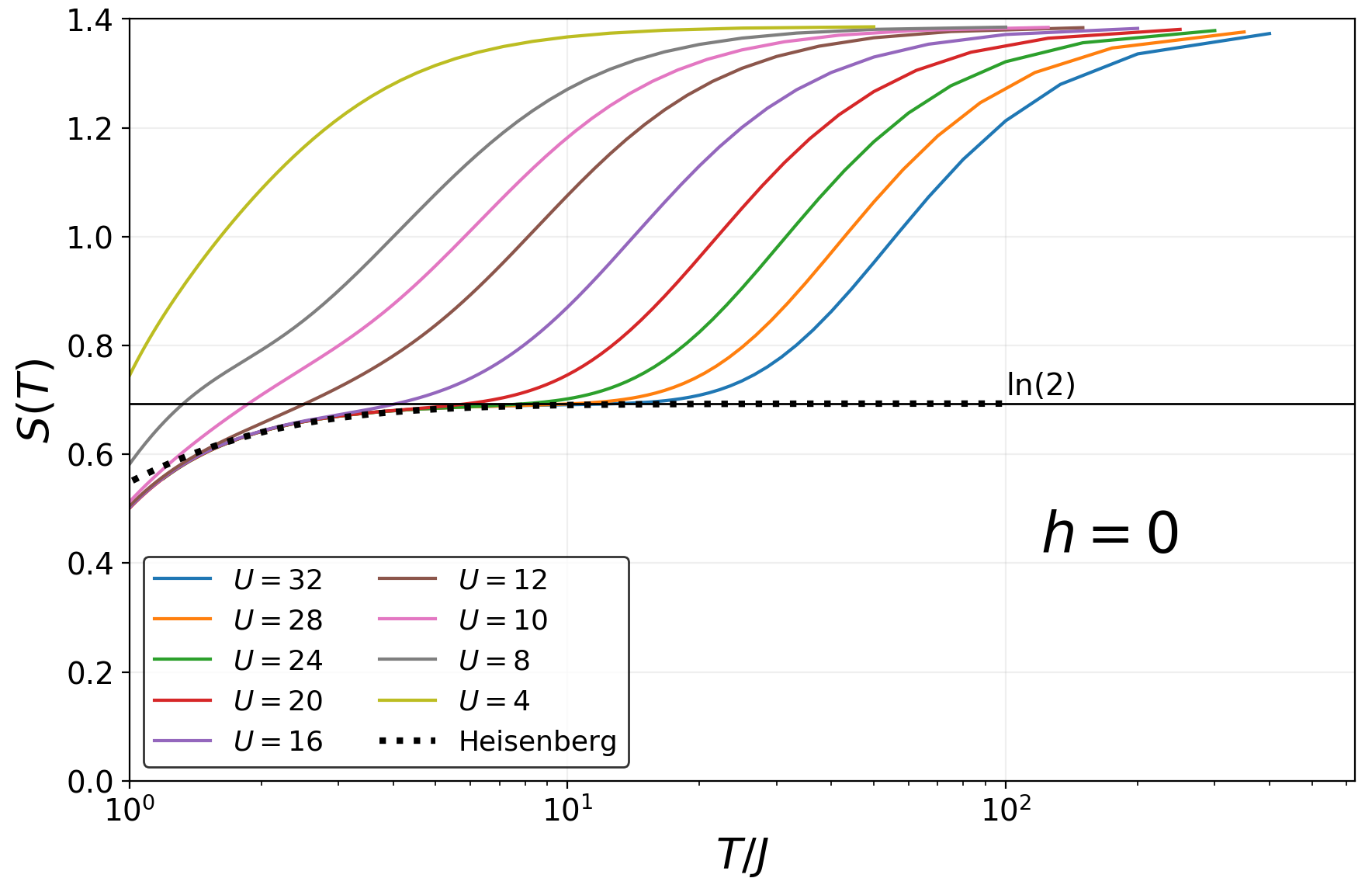}
    \caption{The entropy $S(T)$ at $h=0$ is shown for different values of $U$, along with the triangular-lattice Heisenberg result (dotted line). We observe an entropy plateau at $\ln(2)$ for large $U$, and a high $T$ maximum at $\ln(4)$ as expected. The deviation of $S(T)$ from the Heisenberg limit becomes apparent as $U$ decreases.}
    \label{S_vs_T}
\end{figure}

In Fig.~\ref{S_vs_T}, we plot the entropy $S(T)$ as a function of temperature for several values of $U$ from $U=4$ to $U=32$. One can see that for $U/t>16$, the entropy develops a plateau at a value of $R\ln{2}$ marking the onset of Mott behavior and below that temperature the entropy function is well described by that of the Heisenberg model (a $9^{th}$ order NLC calculation is shown). For intermediate $U/t$ values, there is no entropy plateau but instead the crossover to Heisenberg behavior is obtained at entropy values that are successively smaller than $R\ln{2}$ as $U/t$
is lowered. This crossover happens at temperatures of only a few times $J$ and thus there is no 
high-temperature Heisenberg limit. For $U/t<8$, the entropy is much larger than $R\ln{2}$ at $T=2J$ but 
drops rapidly below that temperature presumably reflecting a crossover to the metallic Fermi liquid regime.

\begin{figure}[t!]
    \centering
    \includegraphics[width=\columnwidth]{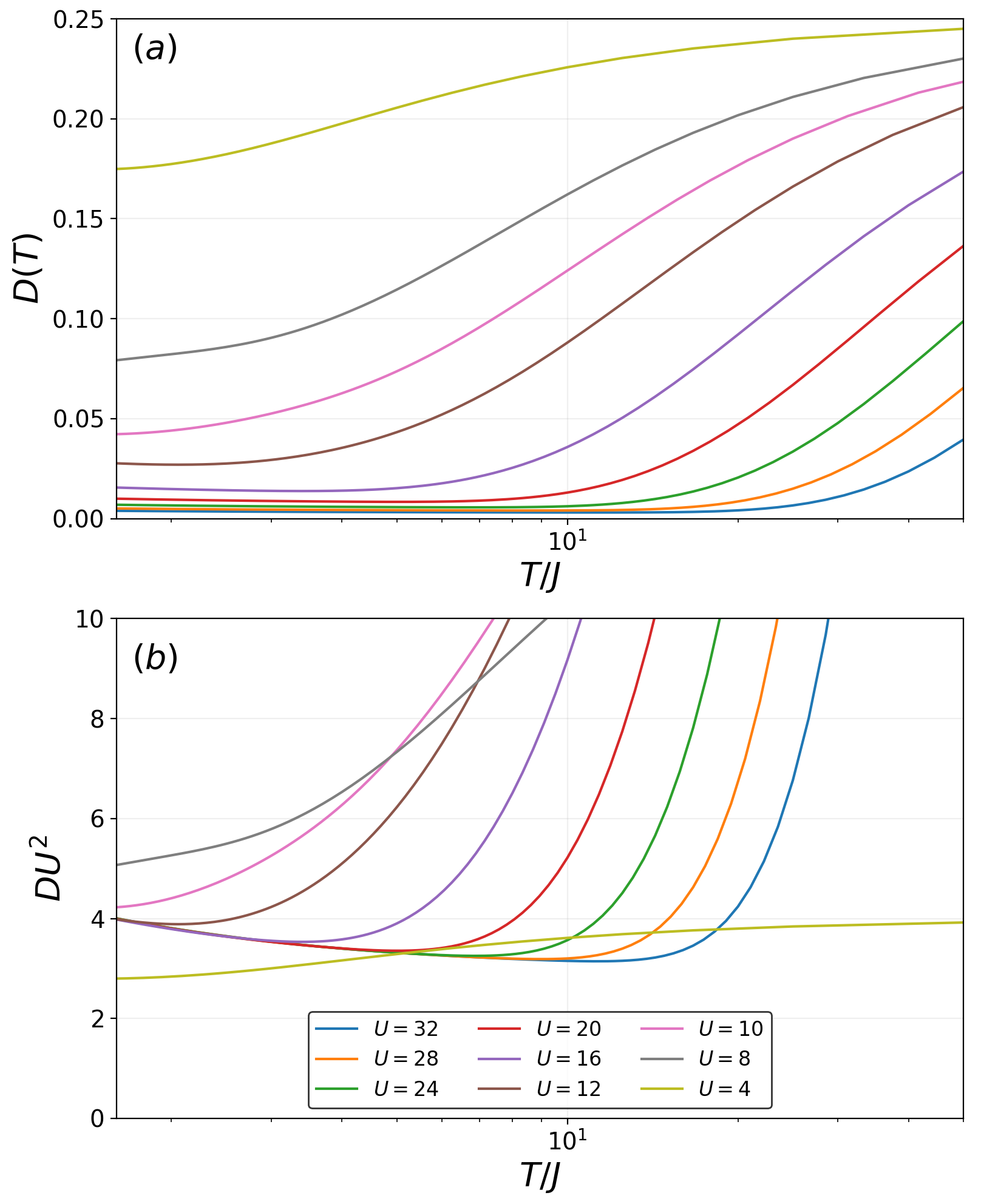}
    \caption{(a) The double occupancy factor $D(T)$ at $h=0$ is shown for different values of $U$. We observe $D(T)$ goes to zero at low $T$ more rapidly at large $U$, and reaches a maximum value of $1/4$ at high temperature as expected. (b) Plot of $DU^2$ as a function of $T$ for different values of $U$, illustrating an upturn for $U\gtrsim12$ at low $T$.}
    \label{D_vs_T}
    %\label{fig:my_label}
\end{figure}

In Fig.~\ref{D_vs_T}, we show the variation of the double occupancy factor $D$ as a function of temperature. In Fig.~\ref{D_vs_T}(a) we plot $D$ whereas in Fig.~\ref{D_vs_T}(b) we plot $DU^2$ versus $T$. In the large $U$ limit double occupancy can
be shown by perturbation theory to go as $1/U^2$. This virtual double occupancy is intimately related to antiferromagnetism and increases as temperature is lowered indicating a growth in antiferromagnetic
configurations compared to ferromagnetic ones. We find that for $U/t>16$, all the plots share the large $U$ behavior at low temperatures. Down to the lowest temperature studied, $U/t=8$ still has substantial double occupancy and it is still slowly decreasing with temperature. The behavior at $U/t=4$ is clearly very different. In this case, the substantial
double occupancy presumably remains in the Fermi liquid regime and is only enhanced by temperature.

\begin{figure}
    \centering
    \includegraphics[width=\columnwidth]{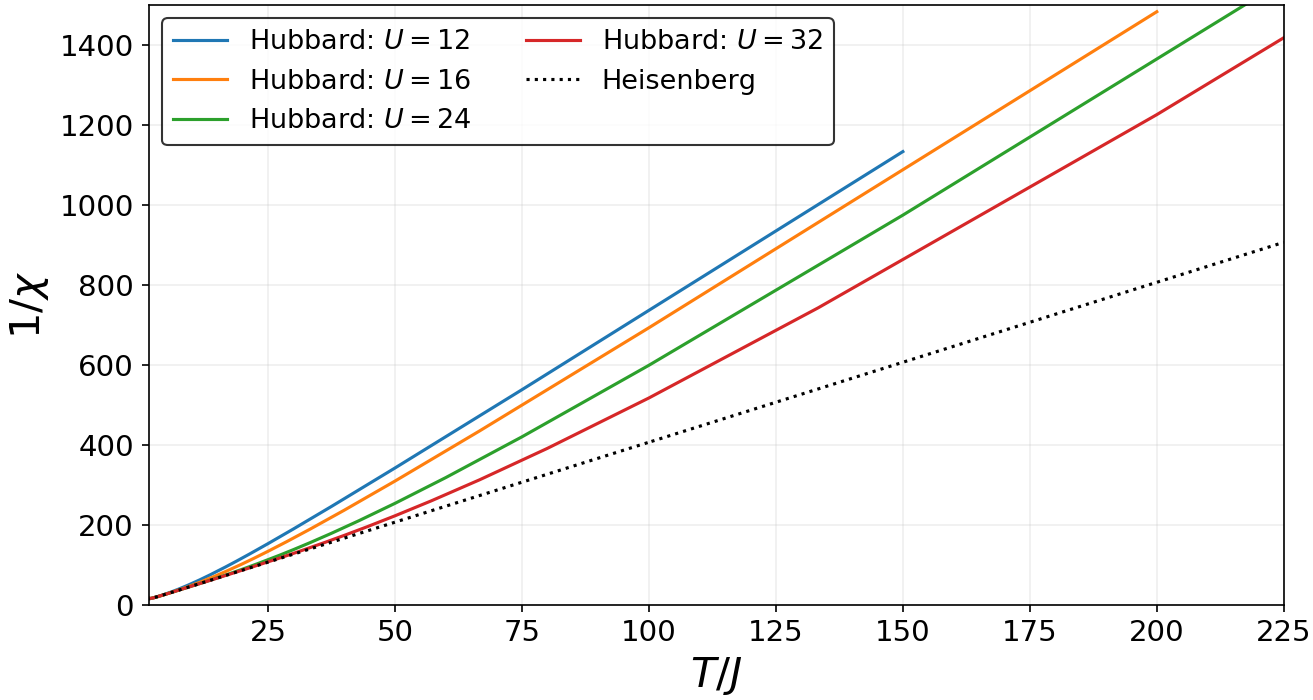}
    \caption{The zero-field inverse magnetic susceptibility $1/\chi$ for the Hubbard model is shown as a function of temperature $T/J$, for $U=12, 16, 24$, and $32$. The Heisenberg model result is shown as a dotted black line.}
    \label{Chi_vs_T}
\end{figure}

In Fig.~\ref{Chi_vs_T}, we show the inverse susceptibility as a function of temperature. The dotted line is the Curie-Weiss result for the Heisenberg model from a $9^{th}$ order NLC calculation. One can see that the susceptibility matches on to the 
Heisenberg limit at temperatures much smaller than $U$. At temperatures above $U$ double occupancy reduces
the effective Curie constant by a factor of $2$. This increases the slope of the inverse susceptibility
plots by a factor of $2$. For a material well described by the Hubbard model, measurements of such a crossover can be a good way to determine the $U/t$ value.

\begin{figure}
    \centering
    \includegraphics[width=\columnwidth]{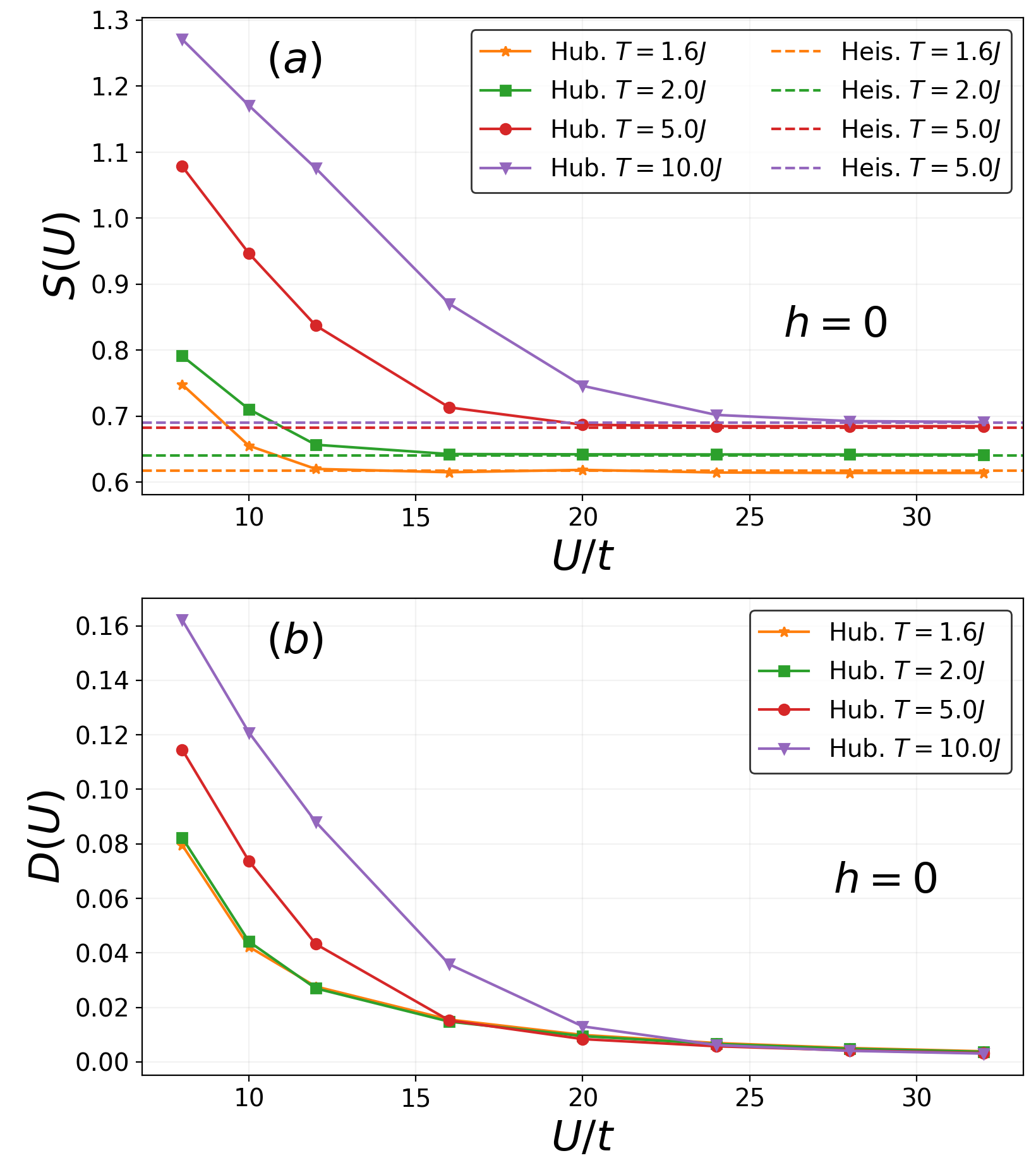}
    \caption{We show (a) entropy $S(U)$ and (b) double occupancy $D(U)$ as a function of $U$, in zero magnetic field. In panel (a) the corresponding entropy values in the Heisenberg model ($9^{th}$ order NLC) are shown as dashed lines. If $U$ is sufficiently large, $S$ approaches the Heisenberg result, with the required $U$ value increasing with temperature.} %$DU^2$ shows a peak around $U \sim 12$ which vanishes as $T$ is lowered. At low $T$ (e.g.~$T=1.6$), there is an upturn in $DU^2$ for $U \lesssim 12$.}
    \label{SD_vs_U}
\end{figure}

In Fig.~\ref{SD_vs_U}, we show the variation of entropy and double occupancy as a function of $U/t$. 
In Fig.~\ref{SD_vs_U}(a) the entropy is plotted as a function of $U/t$ for various temperature values.
In Fig.~\ref{SD_vs_U}(b) the double occupancy is plotted as a function of $U/t$ for various temperature values.
At large $U/t$, the entropy goes smoothly to the values obtained in the Heisenberg limit. As the temperature is lowered, the metal insulator transition should show up as a sharp increase in double occupancy. 
The lowest temperatures where our results are well converged is around $T=1.5 J$. 
At these temperatures one can see a change in
behavior around $U/t=10$, which is a rough indication of the metal-insulator transition, which strictly
occurs only at $T=0$. All properties are smooth at finite temperatures and one cannot identify any
phase transitions more precisely at these temperatures.

\section{IV. Thermodynamic properties in a magnetic field}
\label{sec:resultsB}
In this section, we discuss how the intermediate temperature properties of the Hubbard model evolve with a magnetic field. We consider a Zeeman-field
only. In a quasi-two dimensional system, an in-plane magnetic field only couples to the spin degrees of freedom and acts as a Zeeman field. The application of such a field lowers double occupancy, lowers the entropy and causes the
system to develop a magnetization. Thus we present results for $S(T,h)$, $M(T,h)$ and $D(T,h)$ for several
different parameter ratios $U/t$. All plots are restricted
to temperatures $T>2J$, which is where convergence is best.

\begin{figure}[h]
    \centering
    \includegraphics[width=\columnwidth]{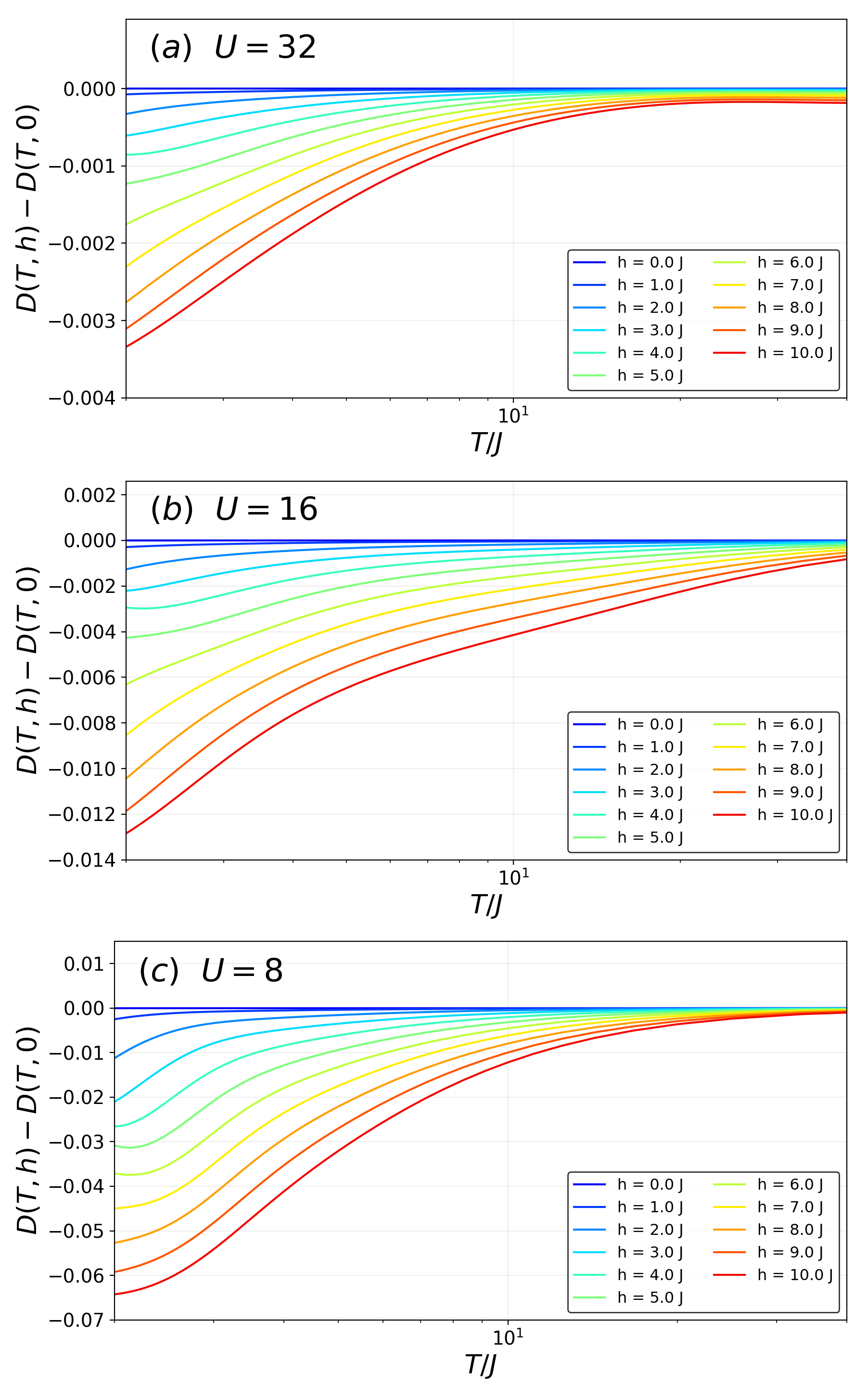}
    \caption{The double occupancy factor difference $D(T,h)-D(T,0)$ is shown for various magnetic field values $h$, for the Hubbard model at (a) $U=32$, (b) $U=16$, and (c) $U=8$. }
    \label{Dh_vs_T}
\end{figure}

In Fig.~\ref{Dh_vs_T}, we show how that the double occupancy is reduced by the application of the field. We plot the difference in the double occupancy parameter in a field and in zero field at a given temperature. A negative value shows it to be reduced. This reduction at large $U$ is related to the fact that the field favors configurations where spins are aligned with the field. Such configurations have reduced virtual mixing with doubly occupied states. 
In zero-field, there was an upturn in $D(T)$ at low temperatures. It goes away at large $h$, e.g.~at $U=32$, $D(T)$ decreases monotonically with temperature for $h/J \gtrsim 5$. As $U$ is decreased, the field has a greater suppression of the double occupancy. This is simply related to the fact that there is more double occupancy in the system in zero field.

\begin{figure}[t!]
    \centering
    \includegraphics[width=\columnwidth]{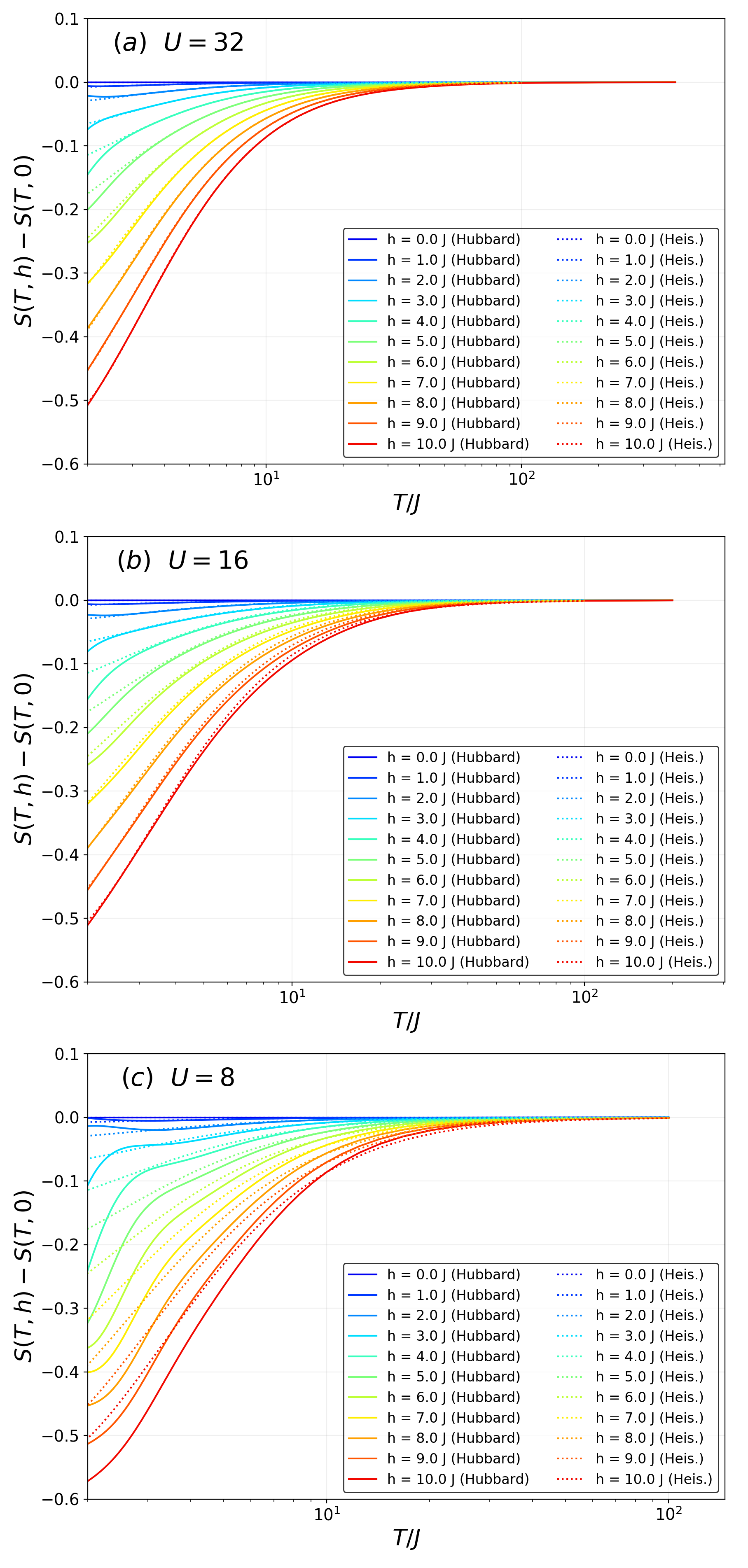}
    \caption{The entropy difference $S(T,h)-S(T,0)$ is shown for various magnetic field values $h$, for the Hubbard model at (a) $U=32$, (b) $U=16$, and (c) $U=8$. }
    \label{Sh_vs_T}
\end{figure}

In Fig.~\ref{Sh_vs_T}, we show how the entropy function is reduced by the application of the field. We plot the difference in the entropy in a field and in zero field at a given temperature. In each panel the Heisenberg result ($9^{th}$ order NLC) is shown for comparison as a dotted line. Qualitatively the results are similar for different U values. At $U=32$, the Hubbard results agree well with the Heisenberg data. Only for $U=8$, we observe significant deviations from the Heisenberg limit with greater deviation occurring at larger fields.

\begin{figure}[h]
    \centering
    \includegraphics[width=\columnwidth]{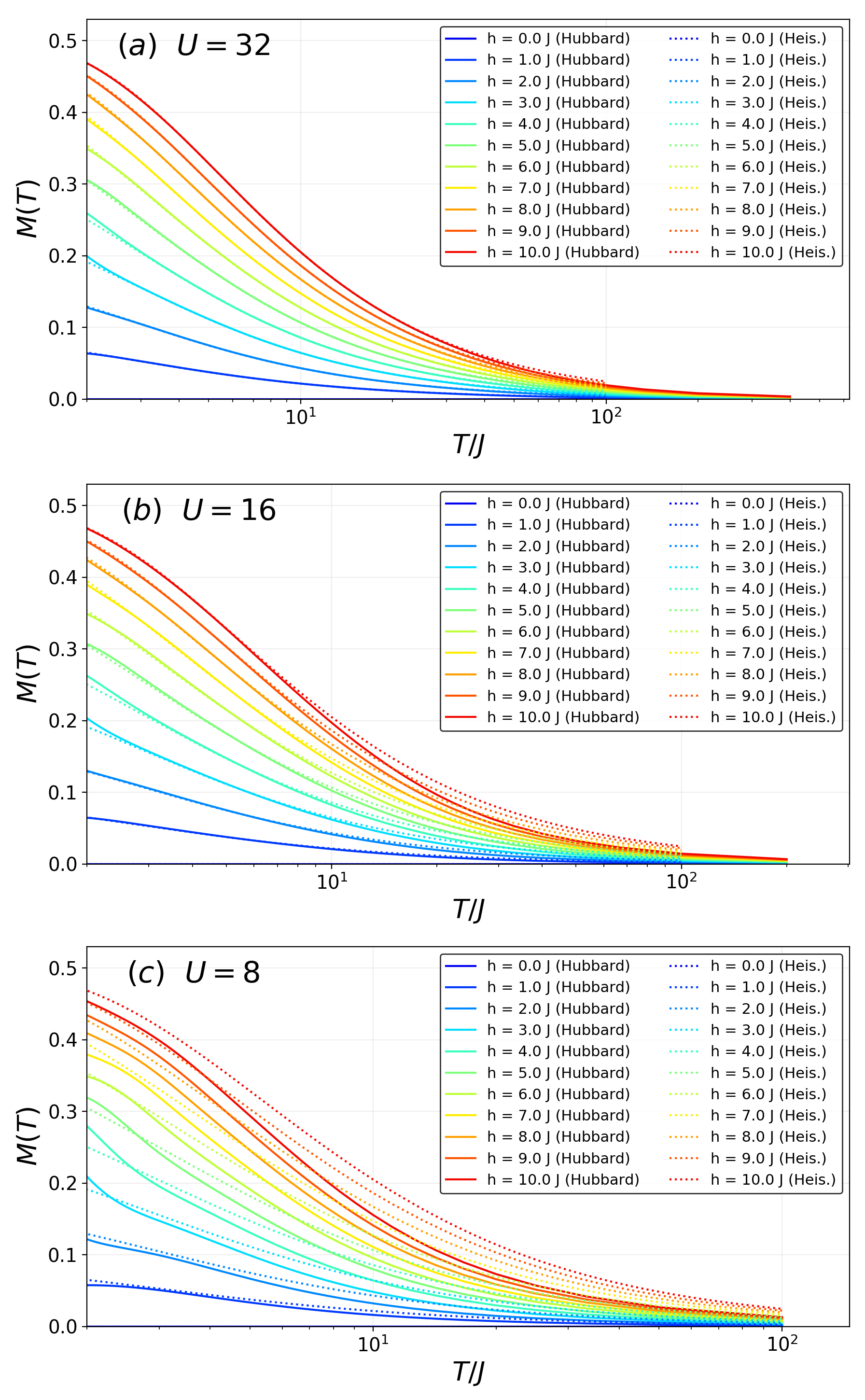}
    \caption{The magnetization $M(T)$ is shown for various magnetic field values $h$, for the Hubbard model at (a) $U=32$, (b) $U=16$, and (c) $U=8$. }
    \label{M_vs_T}
\end{figure}

In Fig.~\ref{M_vs_T}, we show the magnetization as a function of the field. Here too, in each panel the Heisenberg result ($9^{th}$ order NLC) is shown as a dotted line. At $U=32$, the Hubbard results agree well with the Heisenberg data. We observe significant deviation from the Heisenberg limit by $U=8$, again with greater deviation at larger fields.

\begin{figure}[t!]
    \centering
    \includegraphics[width=\columnwidth]{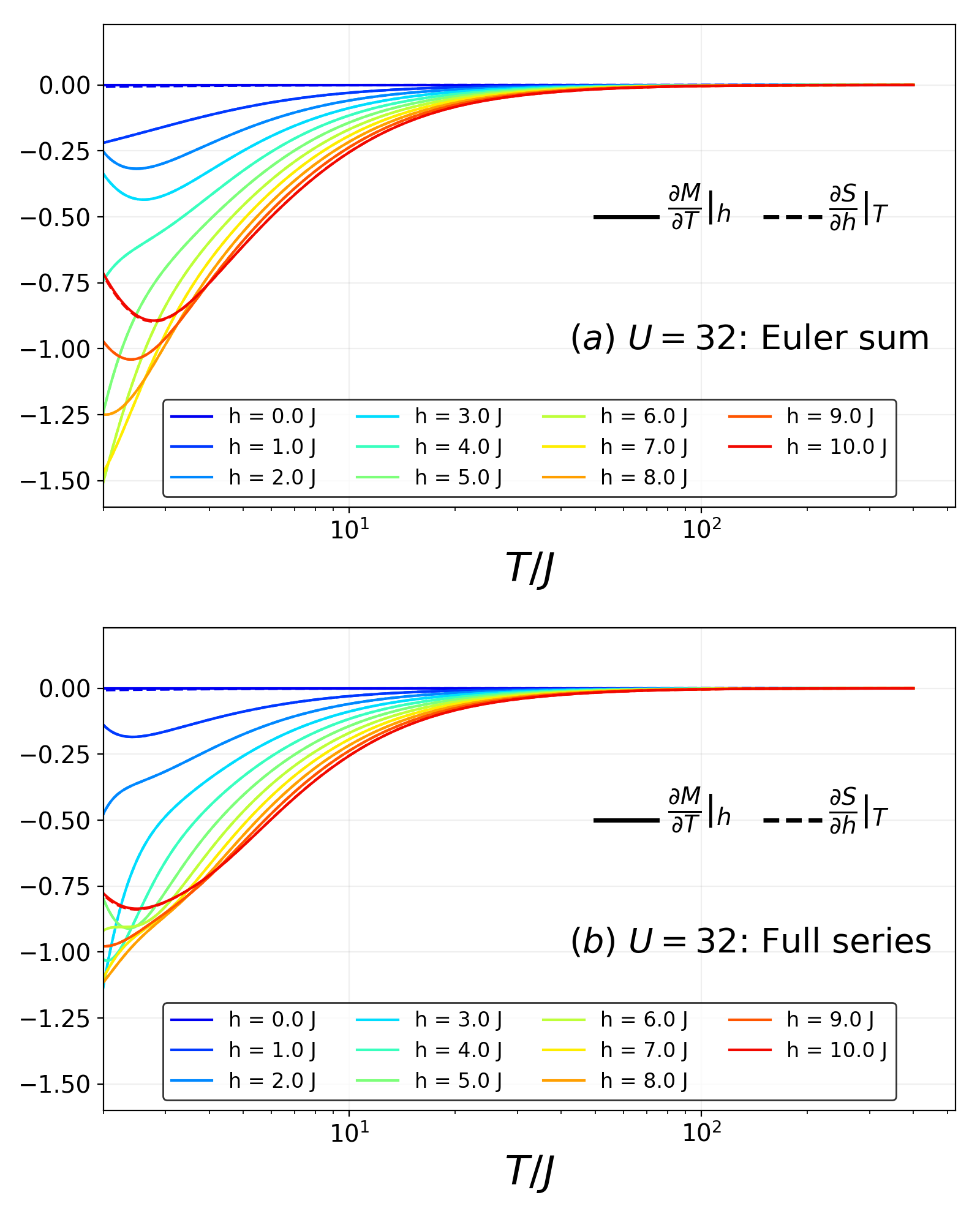}
    \caption{The partial derivatives $(\frac{\partial S}{\partial h})_T$ (dashed lines) and $(\frac{\partial M}{\partial T})_h$ (solid lines) are calculated separately and shown as a function of temperature $T/J$, for several magnetic field values from $h=0$ to $h=10 J$. }
    \label{Maxwell}
\end{figure}

In Fig.~\ref{Maxwell}, The partial derivatives $(\frac{\partial S}{\partial h})_T$ (dashed lines) and $(\frac{\partial M}{\partial T})_h$ (solid lines) are shown as a function of temperature $T/J$, for several magnetic field values from $h=0$ to $h=10 J$. Here we fix $U=32$ and plot for each derivative (a) an $8^{th}$ order Euler sum \cite{Press1989} and (b) the direct sum of the $8^{th}$ order expansion. At large $h$, the magnetization saturates at $1/2$ at low temperatures, thus we observe the derivatives curving upwards towards zero. Maxwell's relation $(\frac{\partial S}{\partial h})_T = (\frac{\partial M}{\partial T})_h$ are well satisfied. The comparison between results obtained from a direct sums of the series in (b) and Euler sums in (a) shows that the convergence of the direct sum starts to break down a bit above $T/J=2$ and it becomes important to employ a convergence method such as Euler summation. The Euler summation improves the convergence in both NLC expansions and series expansions as it eliminates any alternating terms in the series \cite{Hayre2013} however the convergence cannot be pushed too far down in temperature in a reliable way. Hence for field dependent quantities, we mostly show data for $T/J>2$.

\begin{figure}[t!]
    \centering
    \includegraphics[width=\columnwidth]{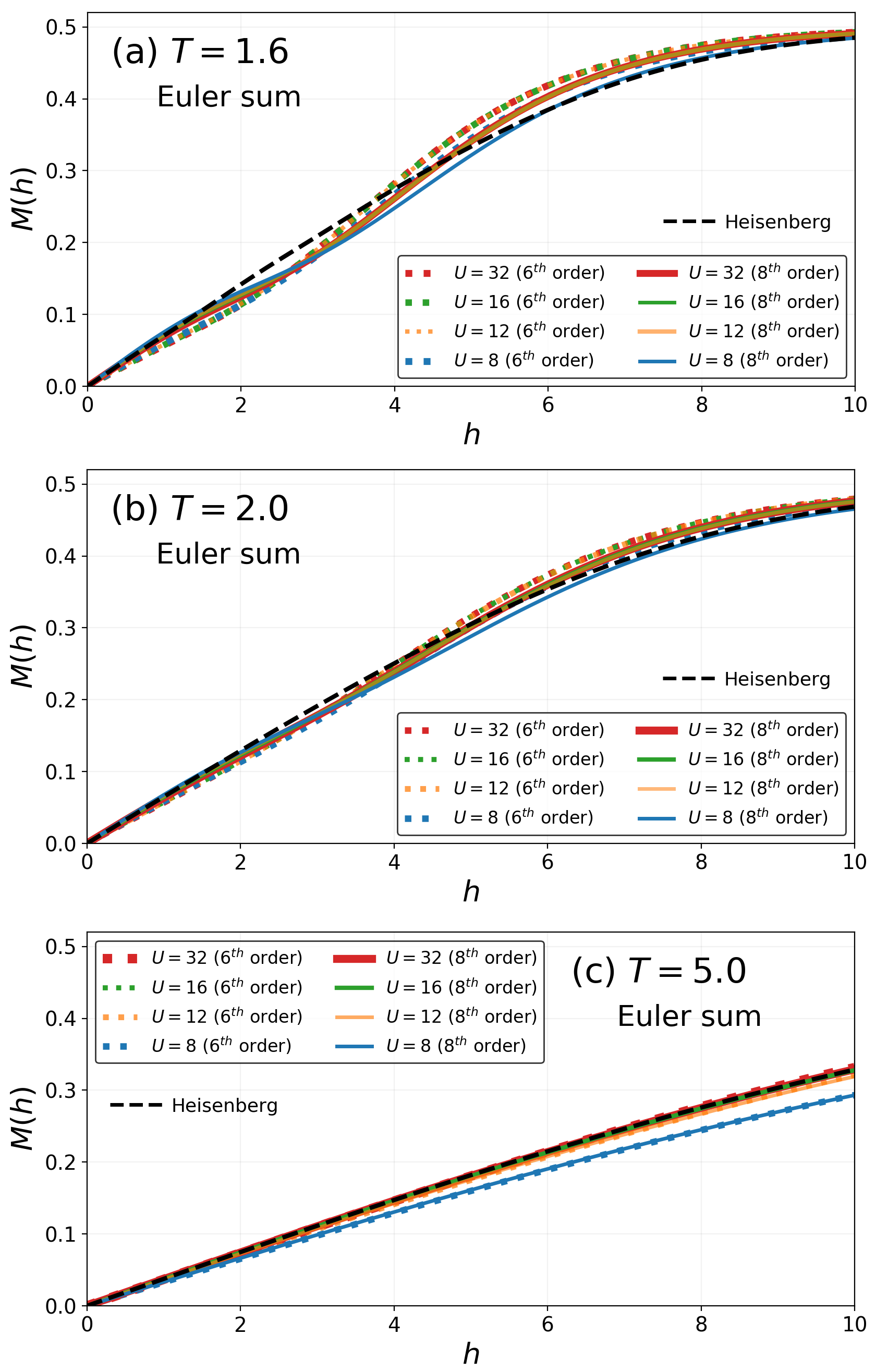}
    \caption{The magnetization $M(h)$ in the triangular lattice Hubbard model at fixed temperatures (a) $T/J=1.6$, (b) $2.0$, and (c) $5.0$ is shown for various values of $U$. In each panel we show an Euler sum for $M(h)$, and compare $6^{th}$ and $8^{th}$ order series. In each plot the corresponding $M(h)$ curve in the triangular lattice Heisenberg model is shown for reference (black dashed line).}
    \label{MvsH}
\end{figure}

\begin{figure}[t!]
    \centering
    \includegraphics[width=\columnwidth]{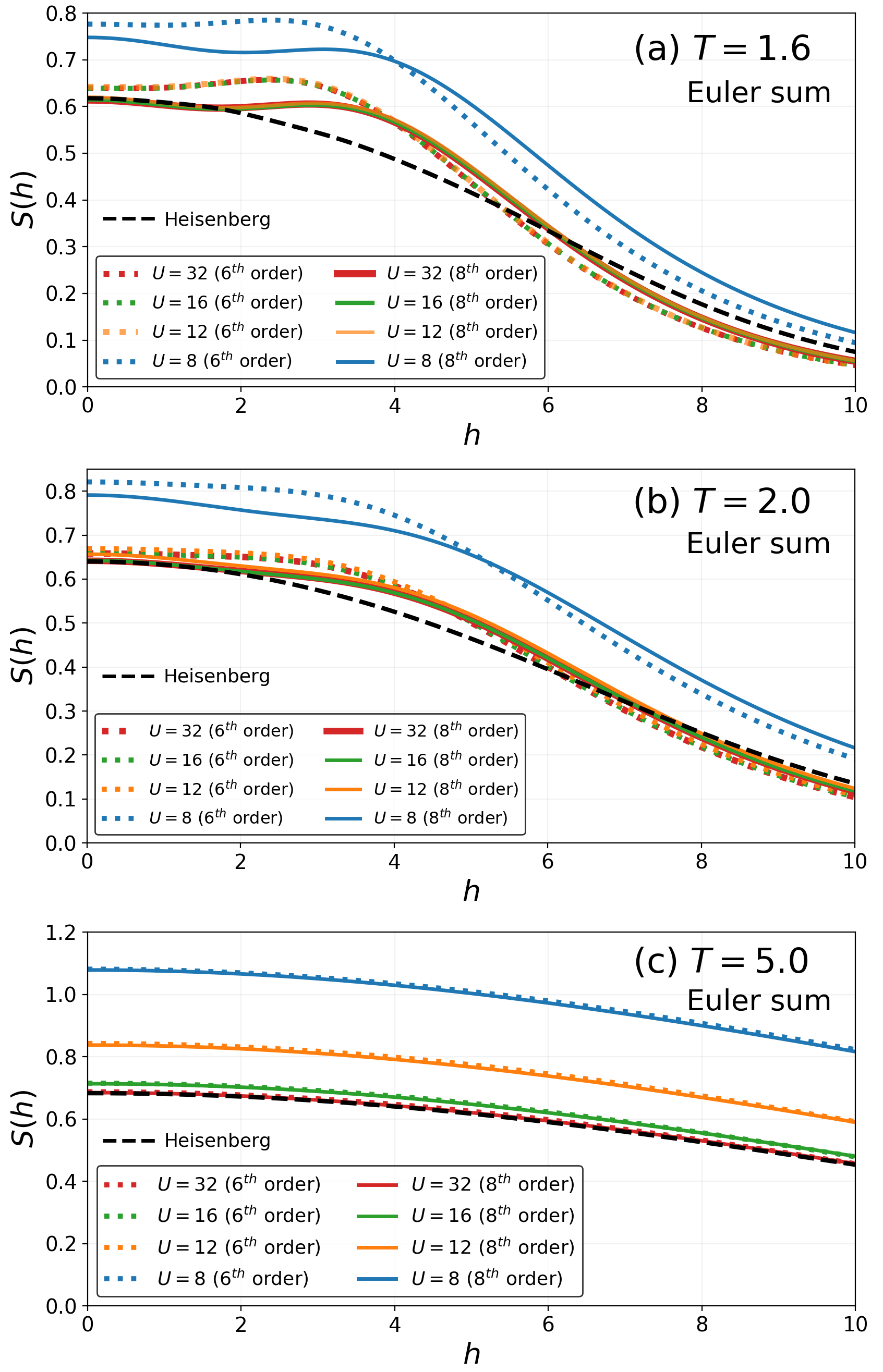}
    \caption{The entropy $S(h)$ in the triangular lattice Hubbard model at fixed temperatures (a) $T/J=1.6$, (b) $2.0$, and (c) $5.0$ is shown for various values of $U$. In each panel we show an Euler sum for $S(h)$, and compare $6^{th}$ and $8^{th}$ order series. In each plot the corresponding $S(h)$ curve in the triangular lattice Heisenberg model is shown for reference (black dashed line).}
    \label{SvsH}
\end{figure}

In Fig.~\ref{MvsH} and Fig.~\ref{SvsH}, we show plots of magnetization and entropy as a function of magnetic field for a few select values of temperature and several values of $U$. At $T=5J$, the field dependence is featureless and the magnetization is converging well towards the Heisenberg limit at large $U$. Here we show Heisenberg model results obtained from a $12^{th}$ order NLC calculation. However, there is still significant additional entropy associated with double occupancy for $U=32$. The excess entropy at $T=2J$ is limited to an intermediate window in the magnetic field for $U=32$. At $T=1.6 J,$ the $M$ vs $h$ curve begins to develop an inflection and at the same time the entropy begins to develop a minima at $h=2J$ and there is even more extra entropy at intermediate fields. While the convergence is starting to break down at these temperatures, we believe that these are an early indication of the magnetization plateaus arising at lower temperatures which may be magnified by the finite $U$ through the frustrating interactions generated in higher orders. This question of precursor effects of magnetization plateaus in magnetization and entropy at temperatures of order $J$ deserve further attention.

%\begin{figure*}[t!]
%    \centering
%    \includegraphics[width=\textwidth]{M(h)_Euler_Sum_Comparison.png}
%    \caption{The magnetization $M(h)$ in the triangular lattice Hubbard model at fixed temperatures $T/J=1.6$, $2.0$, and $5.0$ is shown for various values of $U$. In plots (a), (c), and (e) we show the full expansion of $M(h)$ to both $6^{th}$ order (dotted line) and $8^{th}$ order (solid line). In plots (b), (d), and (f) we show an Euler sum for $M(h)$, and compare $6^{th}$ and $8^{th}$ order series. In each plot the corresponding $M(h)$ curve in the triangular lattice Heisenberg model is shown for reference (black dashed line).}
    %\label{fig:my_label}
%\end{figure*}

%\begin{figure*}[t!]
%    \centering
%    \includegraphics[width=\textwidth]{S(h)_Euler_Sum_Comparison.png}
%    \caption{The entropy $S(h)$ in the triangular lattice Hubbard model at fixed temperatures $T/J=1.6$, $2.0$, and $5.0$ is shown for various values of $U$. In plots (a), (c), and (e) we show the full expansion of $S(h)$ to both $6^{th}$ order (dotted line) and $8^{th}$ order (solid line). In plots (b), (d), and (f) we show an Euler sum for $S(h)$, and compare $6^{th}$ and $8^{th}$ order series. In each plot the corresponding $S(h)$ curve in the triangular lattice Heisenberg model is shown for reference (black dashed line).}
    %\label{fig:my_label}
%\end{figure*}

%\begin{figure*}[t!]
%    \includegraphics[width=\textwidth]{M(h)_U32_U16_Euler_Sum_Comparison.png}
%    \caption{$M(h)$ for both the Hubbard model (color curves) and Heisenberg model (greyscale) for temperatures $T/J = 1.6$, $2.0$, $5.0$, and $10.0$. In panels (a) and (b) we show results for $U=32$, displaying the full series expansion for $M(h)$ and an Euler sum, respectively. In panels (c) and (d) we show results for $U=16$, displaying the full series expansion for $M(h)$ and an order Euler sum, respectively. For each plot, expansions for $M(h)$ are shown to both $6^{th}$ order (dotted line) and $8^{th}$ order (solid line). At low temperatures, we find convergence is generally worse for intermediate fields $h \sim 2J$.}
    
        %\label{fig:my_label}
%\end{figure*}

\section{V. Comparison with LCSO experimental data}
\label{sec:LCSO}
In this section, we present comparisons of the experimental data \cite{Yang2022} for the material 
 Lu$_3$Cu$_2$Sb$_3$O$_{14} $ (LCSO) with those obtained here for the Hubbard model. We have done extensive comparisons of the magnetization data as a function of temperature and magnetic field. First, we use the high-temperature data to estimate the $g$-factor by comparing magnetization $M(T)$ with a second order NLC calculation \cite{Rigol2006,Rigol2007}, which should be highly accurate above $T=10J$. It is known that there are two types of spins in the system, where the exchange constant of one type is larger than the other. For simplicity we assume that one set of spins are coupled much more strongly than the other. At high temperatures only the average exchange that sets the Curie-Weiss constant matters. 
 
\begin{figure}[t!]
    \centering
    \includegraphics[width=\columnwidth]{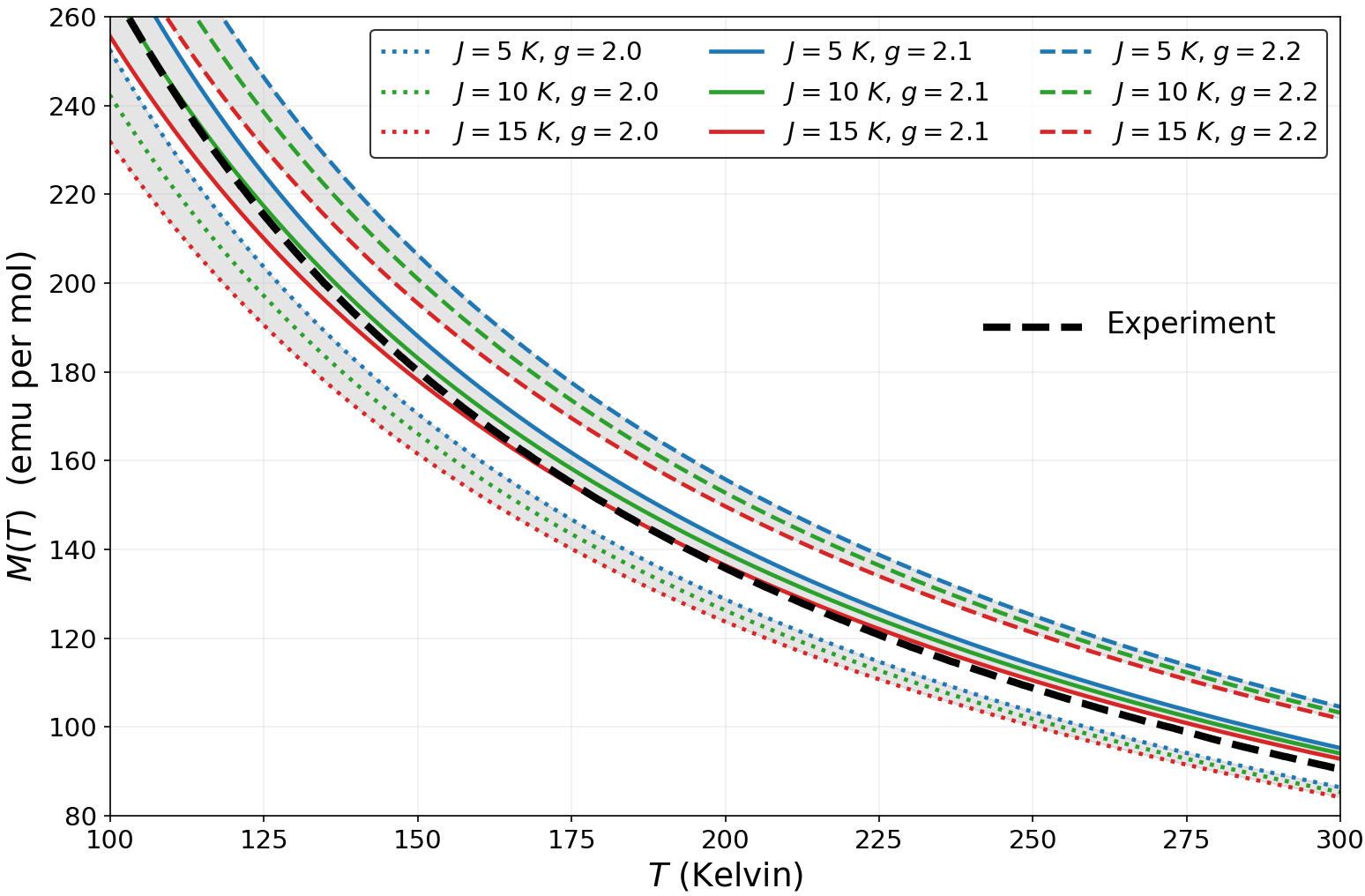}
    \caption{A fit of the very high temperature magnetization data ($T>100K$) at a field of $7$ Tesla for LCSO \cite{Yang2022} with a Heisenberg model on the triangular lattice obtained in second order NLC. We show results with half the spins assumed to be free while the other half are exchange coupled with nearest-neighbor coupling $J$. Different $J$ and $g$ values are shown. Although the very high temperature data are better fit by a $g$ value below $2.1$, the latter value is better for fits at lower temperatures.}
    \label{g_plot}
\end{figure}

We can treat the weakly coupled spin as a free spin, whose magnetization we call $M_1$, while the magnetization of the other will be accurately described by an expression given by a second order NLC calculation. This expression is $3M_2 - 5M_1$, where $M_2$ is the magnetization of a two-site Heisenberg model. An overall magnetization $M(T)$ per spin can then be calculated by averaging over the two types of spin, i.e.
\begin{equation}
M(T) = \frac{1}{2}\left(3M_2 - 5M_1 \right) + \frac{M_1}{2}.
\end{equation}
%In Fig.~\ref{g_plot} we show $\frac{1}{2}(3M_2 - 5M_1) + \frac{M_1}{2}$, which is the average of the two-site NLC result for magnetization (where $M_2$ is the result for a two-site Heisenberg model) and the result for a free spin ($M_1$), 
We plot this in Fig.~\ref{g_plot} for several values of $g$: $2.0$, $2.1$, and $2.2$. For each $g$ value we show $M(T)$ for three values of $J$: $5K$, $10K$, and $15K$, and shade the region enclosed by these curves in gray. The experimental result for the material LCSO (at a field $h=7$ Tesla) is shown as a black dashed line. We find that over the temperature range shown, $g\approx2.1$ fits the data best, although we note that at higher temperatures ($T>300K$) a smaller value $2.0 < g < 2.1$ matches the experimental results more closely. However, since $g=2.1$ fits well over a wide temperature range, we fix $g$ at this value in what follows.

\begin{figure*}[b]
    \centering
    \includegraphics[width=1.9\columnwidth]{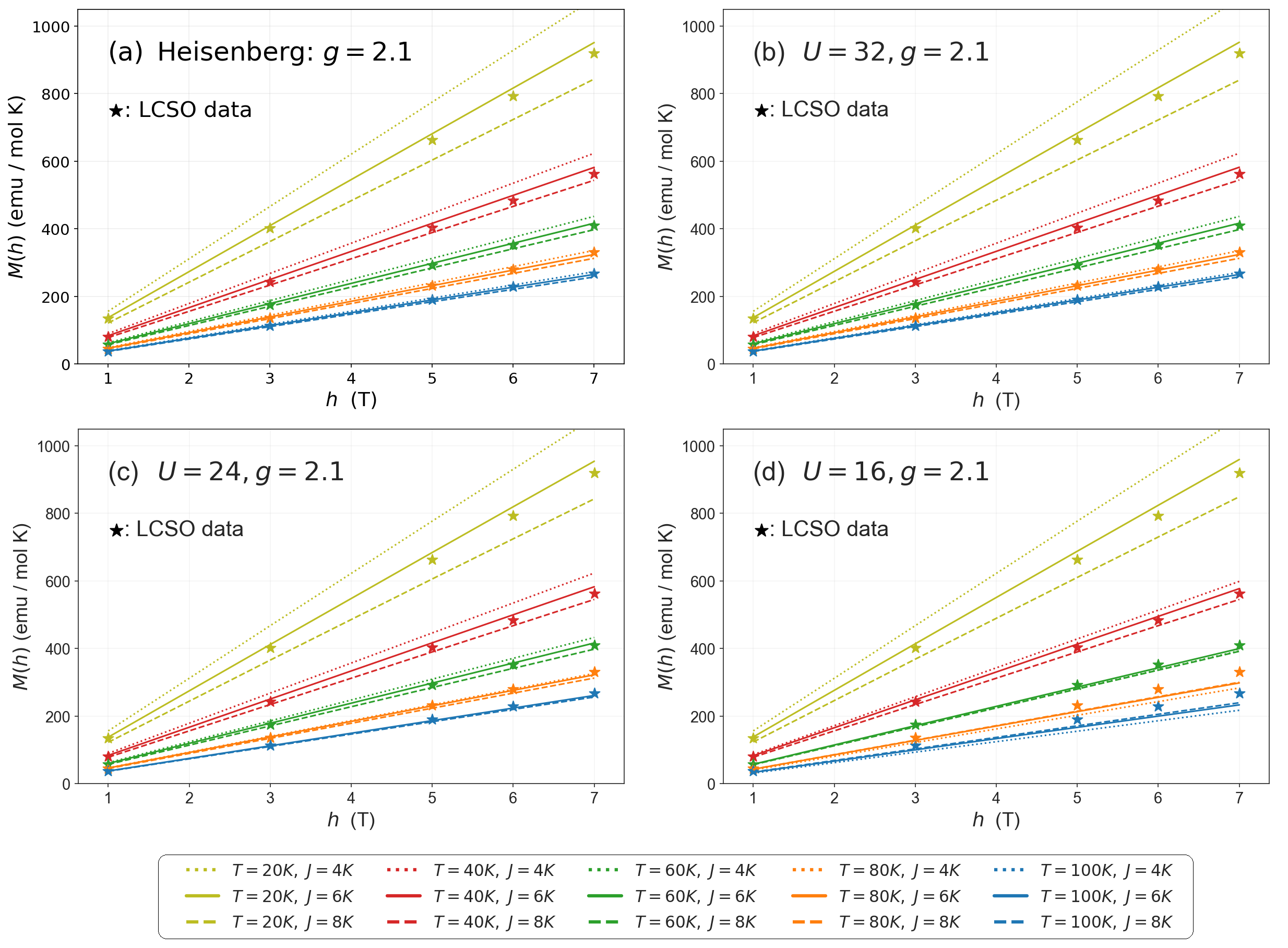}
    \caption{Comparisons of the LCSO experimental data \cite{Yang2022} at select temperatures as a function of magnetic
    field with (a) Heisenberg model and (b)--(d) Hubbard models for different $U/t$ ratios. Best fits are obtained for the Heisenberg or $U=32$ Hubbard model whose results are barely distinguishable from that of the Heisenberg model at fields up to 7 Tesla, which is still in the linear regime. Here all spins are assumed to have the same exchange coupling.}
    \label{experiments}
\end{figure*}

\begin{figure*}[t!]
    \centering
    \includegraphics[width=1.9\columnwidth]{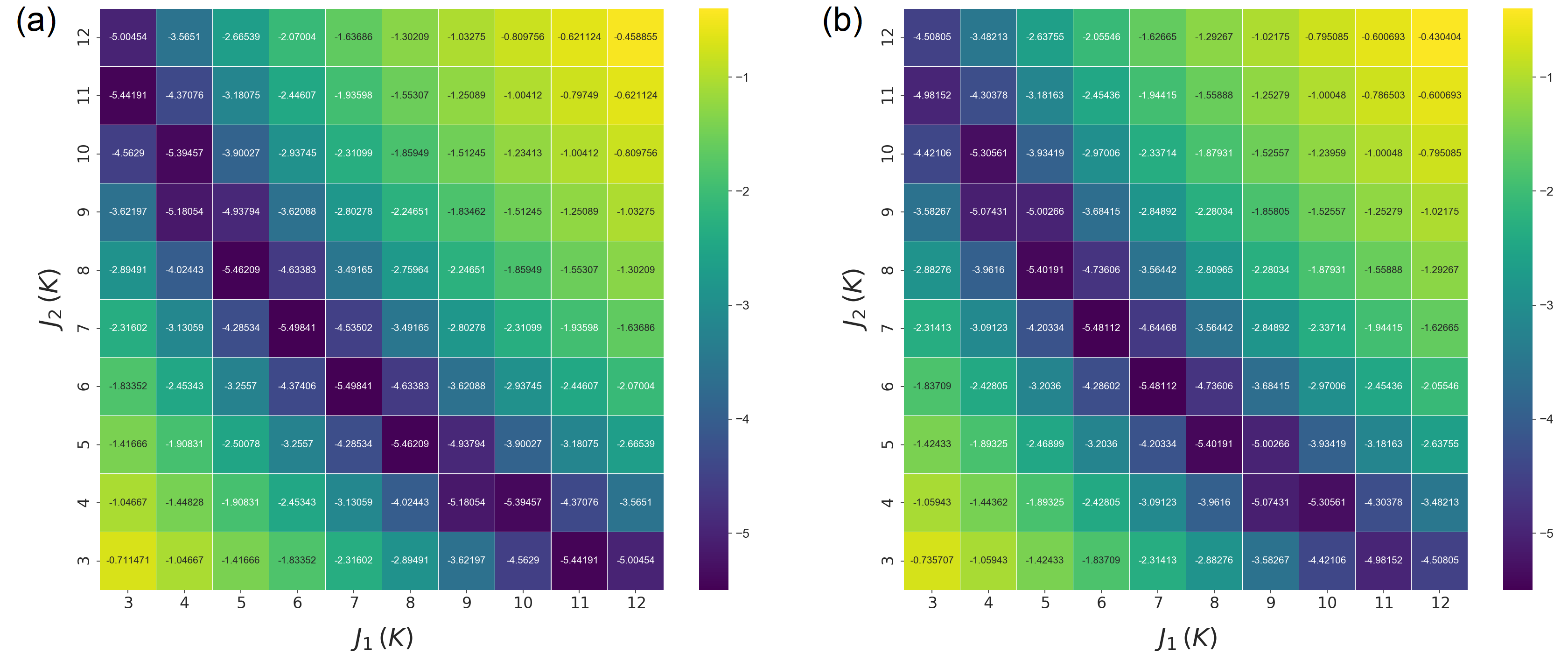}
    \caption{A heat map of the logarithm of relative least square error $\ln(L)$ defined by Eq.~\eqref{L-eqn}. Plots are shown for (a) Heisenberg and (b) $U=32$ Hubbard model. In both cases, we assume the exchange or hopping between the two types of spins are different, giving rise to different exchange constants $J_1$ and $J_2$. The comparison strongly constraints the sum of exchange constants, which also determines the Curie-Weiss constant but not their ratio.}
    \label{heat-map}
\end{figure*}

Next, we determine the optimal value of the exchange parameter $J=4t^2/U$, by comparing both our Hubbard model calculations (at several $U$ values) and $9^{th}$ order Heisenberg NLC results to experimental $M(h)$ data. Here we have assumed that we can represent the system by a single $J$ or single $t/U$ for all spins. We see from Fig.~\ref{experiments} that the fits improve with increasing $U$ and there is not much difference between $U=32$ Hubbard model and the Heisenberg fits. The average $J$, assuming a single exchange constant for all the spins, is greater than $6K$ but less than $8K$. To judge the quality of the fit and allow the interactions for the two types of spins to be different,
we use the quantitative measure of sum of relative least squares. Let the data points be given as $y_i$ and the corresponding calculated values $f_i$. We define the sum of relative least squares as
 \begin{equation}
     L=\sum_i \left(\frac{y_i-f_i}{y_i}\right)^2,
     \label{L-eqn}
 \end{equation}
 %where there are $N$ data points in the fit.
 and plot the logarithm of this quantity to highlight the location of the minima. Heat maps showing $\ln(L)$  for $U=32$ Hubbard and Heisenberg models are shown in Fig.~\ref{heat-map}. Once again, there is not much difference between Heisenberg and $U=32$ Hubbard model fits. We see that the data primarily constrain the average of the exchange constants to be $\frac{J_1+J_2}{2}=6.5 K.$ The ratio of the weaker and the stronger couplings are not well constrained by the behavior at these high temperatures.

 \begin{figure*}[t!]
    \centering
    \includegraphics[width=1.9\columnwidth]{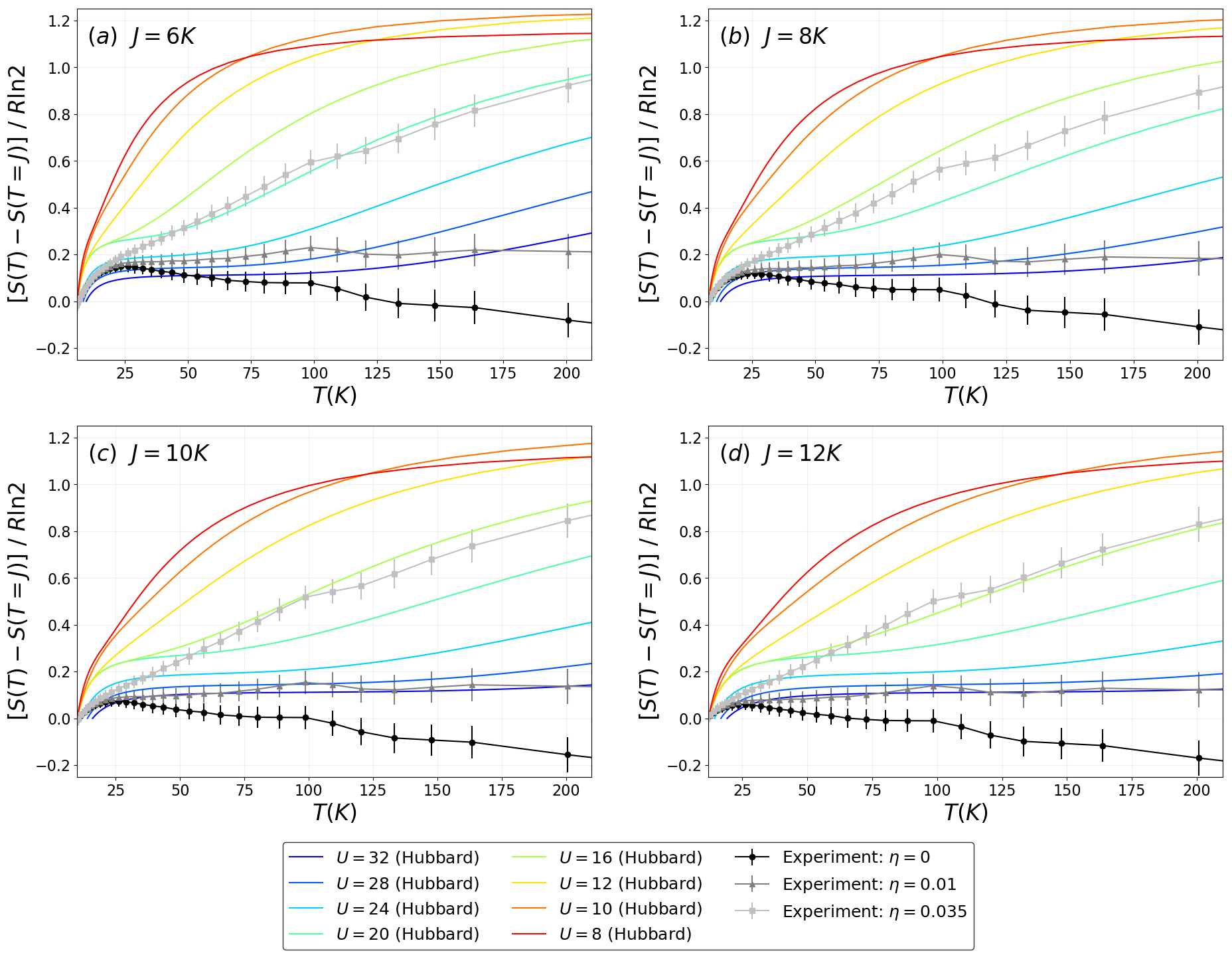}
    \caption{Comparison of the entropy difference $S(T)-S(T=J)$ in zero magnetic field with experimental data for LCSO. In panels (a)--(d) we show the results for a fixed value of the exchange parameter $J$ ranging from $6K$ to $12K$, for several values of $U/t$. Experimental data is shown for three different values of the scaling parameter $\eta$ discussed in Ref.~\cite{Yang2022}. The best fits come from $J$ in the range of 10-12 K and large $U\ge 32$.}
    \label{entropy_experiment}
\end{figure*}

In Fig.~{\ref{entropy_experiment}}, we study the change in entropy $S(T) - S(T=J)$ in zero magnetic field, comparing our Hubbard model results with experimental data for LCSO. Here we have assumed that all of triangular-lattice  spins have the same $U$ and $t$. Results are shown for several values of the exchange parameter $J$, for different $U/t$ ratios ranging from $U/t=8$ to $U/t=32$. As discussed in Ref.~\cite{Yang2022}, the uncertainty in experimental entropy arising in subtraction of lattice entropy from a non-magnetic material can be parameterized by a scale factor $\eta$. We show experimental data for $\eta=0$, $0.01$, and $0.035$ here. We find that the entropy data is best fit by $J \approx 10\textrm{--} 12 K$, for large $U/t$ (i.e.~$U/t = 32$), and a scale factor $\eta=0.01$. Note that the $\eta=0.035$ data is fit well by a smaller $U/t$ value (i.e.~$U/t=20$) with smaller $J\approx 6 K$ but only for temperatures $T \gtrsim 30 K$. However, at temperatures below $T \approx 30 K$, the reduction in entropy is not reflected in the Hubbard model result at these parameters.
 
 A further comparison of the experimental data going down to temperatures below $J$ with a high order NLC calculation for the nearest-neighbor and second-neighbor exchange Heisenberg model is currently in progress and will be discussed elsewhere \cite{Zhang2023}.
 %a free spin. The $g$-factor values range from $2.04$ to $2.1$ with lower values more appropriate if the data over a larger temperature range are considered. 

\section{VI. Discussion and Conclusions} 
\label{sec:conclusions}
In this paper we have studied the magnetization and entropy functions of the triangular-lattice nearest-neighbor Hubbard model at half filling as a function of temperature and magnetic field. These expansions are well converged at temperatures above the hopping parameter $t$. Below temperatures of order $U/2$ there is a crossover from the high temperature regime to the strongly correlated regime for moderate to large values of $U/t$, which is characterized by a suppression of double occupancy and a change in the Curie constant by a factor of $2$. 

Our primary interest in the paper is at temperatures below $t$ and of order $J=\frac{4t^2}{U}$  where the large-$U$ system crosses over to the Heisenberg model. The crossover can be seen quantitatively for all values of $U/t>10$. Only at $U/t$ values below $8$ does the system show behavior that is qualitatively unlike the large $U$ behavior. A strict location of the metal insulator transition cannot be determined at temperatures of order $J$, where our results are convergent, but they are consistent with a transition in the range $8<(U/t)_c<10$.

One of our motivations was to see if finite-$U/t$ value can help explain the unusually small entropy difference between the temperatures of $2J$ and $0.1 J$ seen in the LCSO materials \cite{Yang2022} by already reducing the system entropy at temperatures above $2J$. That is not found to the case. In fact, best fits to the experimental data are obtained by the Hubbard model of $U/t=32$ or larger, whose results in zero-field are barely distinguishable from the Heisenberg model.

We find some evidence that strong magnetic fields with energy scales larger than the exchange constant $J$ can lead to some precursors of low temperature magnetization plateaus with signatures in both magnetization and entropy already at temperatures above $J$. This issue deserves further attention experimentally and theoretically.
 
\section{Acknowledgments} 
We would like to thank Chandra Varma for getting us interested in this problem, Ehsan Khatami and Pranav Seetharaman for help with the Heisenberg model NLC calculations, and Yanxing Yang and Lei Shu for providing us with the experimental data on the LCSO materials. This work was supported in part by the US National Science Foundation grant number DMR-1855111. One of the authors (JO) acknowledges computing support provided by the Australian National Computation Infrastructure (NCI) Program.

% \subsection*{Appendix A: Third order expansion of electron density}
% \setcounter{equation}{0}
% \renewcommand{\theequation}{A\arabic{equation}}

% \begin{figure}[t!]
%     \centering
%     \includegraphics[width=\columnwidth]{Hubbard_Third_Order_Density_Paper.png}
%     \caption{We show the different terms in the third order expansion of the electron density, which sum to give $\rho_3(T)$ (blue curve). Note that $\zeta \equiv e^{\beta \mu}$ is the fugacity of the system. Descending from high $T$, the sum of the components becomes positive (i.e.~$\rho$ will slightly exceed half-filling). Increasing from low $T$, the sum of the components becomes negative (i.e.~$\rho$ will be slightly below half-filling).}
%     %\label{fig:my_label}
% \end{figure}

% \clearpage 
% \newpage 

\bibliography{main}

\end{document}